\documentclass[reprint,prc,showpacs,amsmath,amssymb,superscriptaddress]{revtex4-1}
\usepackage{dcolumn}
\usepackage[dvips]{graphicx}
\usepackage{amsmath}
\usepackage{multirow}
\usepackage{setspace}
\newcommand\T{\rule{0pt}{3.1ex}}
\usepackage{color}
\usepackage{bm}
\usepackage[T1]{fontenc}
\usepackage[latin9]{inputenc}
\usepackage{color}
\usepackage{amssymb}
\usepackage{float}

\providecommand{\tabularnewline}{\\}

\usepackage{array}

\begin{document}

\title{Neutrinoless double electron capture}

\author{J. Kotila}
\email{jenni.kotila@yale.edu}
\affiliation{Center for Theoretical Physics, Sloane Physics Laboratory,
Yale University, New Haven, Connecticut 06520-8120, USA}
\affiliation{Department of Physics, University of Jyv\"askyl\"a, B.O. Box 35, FIN-40014, Jyv\"askyl\"a, Finland}

\author{J.\ Barea}
\email{jbarea@udec.cl}
\affiliation{Departamento de F\'{i}sica, Universidad de Concepci\'{o}n,
 Casilla 160-C, Concepci\'{o}n 4070386, Chile}

\author{F.\ Iachello}
\email{francesco.iachello@yale.edu}
\affiliation{Center for Theoretical Physics, Sloane Physics Laboratory,
 Yale University, New Haven, Connecticut 06520-8120, USA}

\begin{abstract}
Direct determination  of the neutrino mass is at the present time one of the most important aims of experimental and theoretical research in nuclear and particle physics. A possible way of detection is through neutrinoless double electron capture, $0\nu ECEC$. This process can only occur  when the energy of the initial state matches precisely that of the final state. We present here a calculation of prefactors (PF) and nuclear matrix elements (NME) within the framework of the microscopic interacting boson model (IBM-2) for $^{124}$Xe, $^{152}$Gd, $^{156}$Dy, $^{164}$Er, and $^{180}$W. From PF and NME we calculate the expected half-lives and obtain results that are of the same order as those of $ 0\nu \beta^+\beta^+$ decay, but considerably longer than those of $0\nu \beta^-\beta^-$ decay.

\end{abstract}

\maketitle

\section{Introduction}
\begin{figure}[cbt!]
\begin{center}
\includegraphics[width=8.6cm]{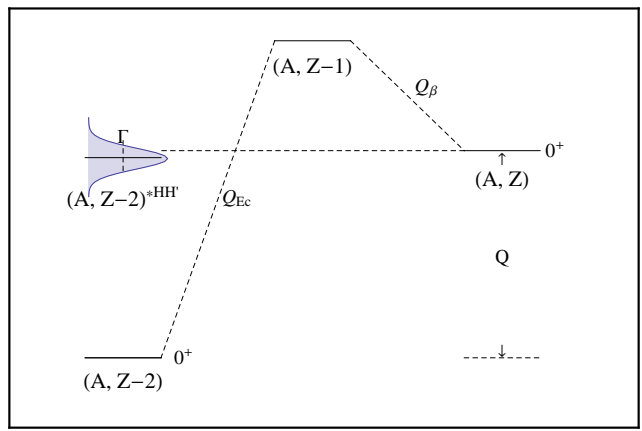} 
\end{center}
\caption{\label{fig1} (Color online) Schematic representation of neutrinoless double electron capture.}
\end{figure}

\begin{figure}[cbt!]
\begin{center}
\includegraphics[width=8.6cm]{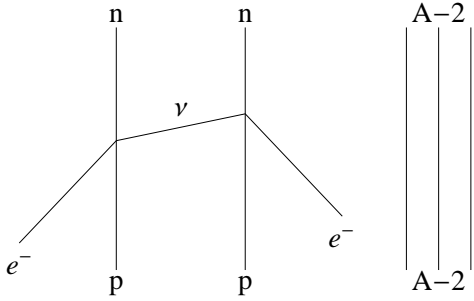} 
\end{center}
\caption{\label{fig2} Neutrinoless double electron capture where $\nu=\bar{\nu}$ can be either light or heavy neutrino.}
\end{figure}

The question whether or not the neutrino is a Majorana particle and, if so, what is its average mass remains one
the of the most fundamental problems in physics today. In a previous series of papers we have calculated phase-space factors (PSF) and nuclear matrix elements (NME) for $0\nu\beta^-\beta^-$, $2\nu\beta^-\beta^-$ processes \cite{bar09,kot12,bar12,bar12b, kot12c,bar12c, fin12, bell13,iac13}, and for $0\nu\beta^+\beta^+$, $0\nu EC\beta^+$, and $2\nu\beta^+\beta^+$, $2\nu EC\beta^+$ and $2\nu ECEC$ processes \cite{kot13,bar13b}. The neutrinoless electron capture $0\nu ECEC$ was not calculated due to the fact that, in general, one cannot conserve energy and momentum in the process
\begin{equation}
(A,Z) + 2e^{-} \rightarrow(A,Z-2).
\end{equation}
However, conservation of energy and momentum can occur in the special case in which the energy of the initial state matches precisely the energy of the final state. This situation was first discussed by Winter \cite{win55} and subsequently elaborated by Bernabeu \textit{et al.} \cite{ber83}, who provided estimates of the inverse life-times with simple NME and PF. The work of \cite{ber83} stimulated many experimental searches and additional theoretical papers \cite{zuj04,luk06,bel09,sim09,rah09,kol10,bel11,kri11,eli11,kol11,gon11,eli11b,eli11d,eli11c,ver11,suh12b,suh12,rod12,fan12,nes12,smo12,dro12,bel13,suh13}. The process was also termed resonant neutrinoless double electron capture, $R0\nu ECEC$. The matching condition can occur either for g.s. to g.s. transitions or for transitions between the g.s. and an excited state in the final nucleus. The precise matching condition is an exceptional circumstance which may or may not occur in practice. A slightly less stringent condition is that the decay occurs through the tail of the width of the atomic initial state as shown schematically in Fig. \ref{fig1}.
 For this process, depicted in Fig. \ref{fig2}, the inverse half-life can be to a good approximation factorized as \cite{ber83}
\begin{equation}
\begin{split}
\left[ \tau_{1/2}^{ECEC}(0^+)\right]^{-1}=&g_A^4 G_{0\nu}^{ECEC}\left|M^{0\nu}_{ECEC}\right|^2\left|f(m_i,U_{ei}) \right|^2\\
&\times\frac{(m_ec^2) \Gamma }{\Delta^2+\Gamma^2 /4},
\end{split}
\end{equation}
where  $G_{0\nu}^{ECEC}$ is a prefactor depending on the probability that a bound electron is found at the nucleus (see Sect. II), $M^{0\nu}_{ECEC}$ is the nuclear matrix element, and $f(m_i,U_{ei})$ contains physics beyond the standard model through the masses $m_i$ and mixing matrix elements $U_{ei}$ of neutrino species. For light neutrino exchange 
\begin{equation}
\begin{split}
f(m_{i},U_{ei})&=\frac{\langle m_{\nu} \rangle}{m_e}, \\
&\langle m_{\nu} \rangle=\sum_{k=light}(U_{ek})^2 m_k,
\end{split}
\end{equation}
while for heavy neutrino exchange 
\begin{equation}
\begin{split}
f(m_{i},U_{ei})&=m_p\langle m_{\nu_h}^{-1} \rangle, \\
& \langle m_{\nu_h}^{-1} \rangle=\sum_{k=heavy}(U_{ek_h})^2\frac{1}{ m_{k_h}}. 
\end{split}
\end{equation}
The last factor, often written as 
\begin{equation}
\frac{(m_ec^2)\Gamma}{\Delta^2+\Gamma^2/4}=(m_ec^2)F
\end{equation}
is the figure of merit for this process. Here  $\Delta=|Q-B_{2h}-E|$ is called the degeneracy parameter, where $\Gamma=\Gamma_{e_1}+\Gamma_{e_2}$  is the two-hole width and $B_{2h}$ is the energy of the double-electron hole in the atomic shell  of the daughter nuclide including binding energies and Coulomb interaction energy.  The importance of the Coulomb interaction energy of the two holes was investigated in Ref.~\cite{kri11}. Obviously the maximum value of $F$ is $F^{\text{max}}=4 / \Gamma$.

\begin{figure}[cbt!]
\begin{center}
\includegraphics[width=8.6cm]{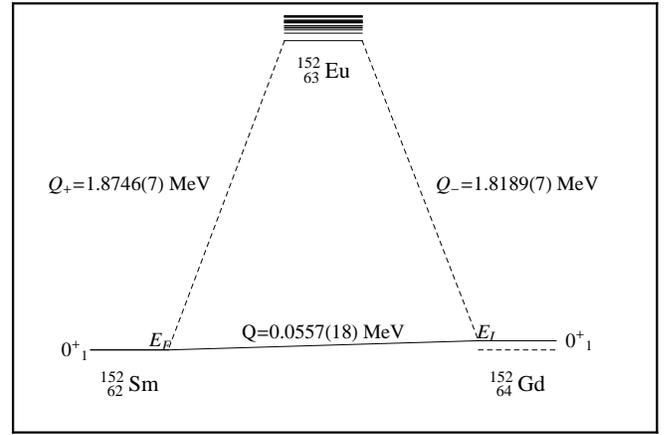} 
\end{center}
\caption{\label{fig3}  Energetics of $^{152}$Gd$\rightarrow ^{152}$Sm double electron capture.}
\end{figure}

Since $\Gamma$ is of the order of eV one needs to find nuclei or nuclear states such that $\Delta$ has the smallest possible value. This requires an accurate measurement of the Q-value. Recent improvements in measurements of mass differences \cite{rah09,kol10,kol11,eli11,gon11,nes12,smo12} have ruled out many of the initial candidates.  In this paper, we report  calculations of the PF and NME of five remaining candidates, $^{124}$Xe$\rightarrow ^{124}$Te$^*$,$^{152}$Gd$\rightarrow ^{152}$Sm, $^{156}$Dy$\rightarrow ^{156}$Gd$^*$, $^{164}$Er$\rightarrow ^{164}$Dy, and $^{180}$W$\rightarrow ^{180}$Hf, where the star denotes excited state. (In addition to these, other candidates remain but  with spin and parity of the final excited state unknown.  Theses cases are not discussed here.)
 The energetics for the 5 cases of interest are shown in Table \ref{table1}. The case $^{152}$Gd$\rightarrow ^{152}$Sm decay is also illustrated in  Fig. \ref{fig3}.

\begin{ruledtabular}
\begin{center}
\begin{table*}[cbt!]
\label{table1}
\caption{Double electron captures considered, the $Q$-value of the decay, energy of the resonant state in the daughter nucleus, capture shells, energy of the double-electron hole, two-hole width, and resonance enhancement factor.}
\begin{tabular}{lcccccccc}
Decay  &$Q$-value(keV)&$E$(keV) &$Q-E$(keV)  &Shells &$B_{2h}$(keV) & $\Delta$(keV)&$\Gamma$(keV)\footnotemark[1] &$(m_ec^2)$F\\
\hline
\T
$_{54}^{124}$Xe$_{70}\rightarrow _{52}^{124}$Te$^*_{72}$ &$2856.73 \pm 0.12$\footnotemark[2]   	&$2790.41\pm 0.09$	&66.32 &$K-K$	&64.457\footnotemark[2] & 1.86 &0.0198	&2.92\\
$_{64}^{152}$Gd$_{88}\rightarrow _{62}^{152}$Sm$_{90}$ &$55.70 \pm 0.18$ \footnotemark[3]  		&$0.0$ 		&55.70 &$K-L_1$ & 54.795\footnotemark[3]&0.91&0.023	&14.38\\
$_{66}^{156}$Dy$_{90}\rightarrow _{64}^{156}$Gd$^*_{92}$ &$2005.95 \pm 0.10$\footnotemark[4] 	  	&$1988.5\pm 0.2$ 	&17.45 &$L_1-L_1$	& 16.914\footnotemark[4]
 & 0.54 &0.0076		&13.52\\
$_{68}^{164}$Er$_{96}\rightarrow _{66}^{164}$Dy$_{98}$ &$25.07 \pm 0.12$\footnotemark[5] 		&$0.0$ 		&25.07 &$L_1-L_1$	& 18.259\footnotemark[5]
 &6.81 &0.0086	&0.095\\
$_{74}^{180}$W$_{106}\rightarrow _{72}^{180}$Hf$_{108}$ &$143.20 \pm 0.27$\footnotemark[6]   		&$0.0$ 	& 143.20	&$K-K$	&131.96\footnotemark[7]\footnotemark[8] & 11.24 &0.072	&0.29\\
\end{tabular}
\footnotetext[1]{Ref.~\cite{ato77}}
\footnotetext[2]{Ref.~\cite{nes12}}
\footnotetext[3]{Ref.~\cite{eli11b}}
\footnotetext[4]{Ref.~\cite{eli11d}}
\footnotetext[5]{Ref.~\cite{eli11c}}
\footnotetext[6]{Ref.~\cite{dro12}}
\footnotetext[7]{Ref.~\cite{lar77}}
\footnotetext[8]{Ref.~\cite{kri11}}

\end{table*}
\end{center}
\end{ruledtabular}

\section{Prefactors}

In the calculation of the prefactor, PF$\equiv G_{0\nu}^{ECEC}$, we follow the theory described in our previous papers \cite{kot12, kot13}, in particular that of \cite{kot13} for electron capture (EC). The captured electrons are described by positive energy Dirac central field bound state wave functions,
\begin{equation}
\psi_{n`\kappa\mu}(\mathbf{r})=\left(
\begin{array}{c}
g^b_{n`,\kappa}(r)\chi_{\kappa}^{\mu}\\
if^b_{n`,\kappa}(r)\chi_{-\kappa}^{\mu},
\end{array}
\right),
\end{equation}
where $n'$ denotes the radial quantum number and the quantum number $\kappa$ is related to the total angular momentum, $j_{\kappa}=|\kappa|-1/2$. 
The bound state wave functions are normalized in the usual way
\begin{equation}
\begin{split}
&\int \psi_{n`\kappa\mu}(\mathbf{r})^{\dagger}\psi_{n`\kappa\mu}(\mathbf{r})\mathrm{d}\mathbf{r}=\\
&\int_0^\infty \left[ \left(g^b_{n`,\kappa}(r)\right)^2+\left(f^b_{n`,\kappa}(r)\right)^2 \right] \mathrm{d}r=1.
\end{split}
\end{equation}
They are calculated numerically by solving the Dirac equation with finite nucleon size and electron screening in the Thomas-Fermi approximation \cite{kot12, kot13}.

For the calculation of electron capture processes the crucial quantity is the probability that an electron is found at the nucleus. This can be expressed in terms of the dimensionless quantity 
\begin{equation}
\begin{split}
{\cal B}_{n',\kappa}^2=&\frac{1}{4\pi(m_ec^2)^3}\left( \frac{\hbar c}{a_0}\right)^3 \left( \frac{a_0}{R}\right) ^2 \\
&\times \left[ \left(g_{n',\kappa}^b(R)\right)^2 +\left(f_{n',\kappa}^b(R)\right)^2 \right],
\end{split}
\end{equation}
where $a_0$ is the Bohr radius $a_0=0.529 \times 10^{-8}$cm and we use for the nuclear radius $R=1.2A^{1/3}$fm. For capture from the $K$-shell $n'=0$, $\kappa=-1$, $1S_{1/2}$ while for capture from the $L_I$-shell $n'=1$, $\kappa=-1$, $2S_{1/2}$.
The PF is given by: 
\begin{equation}
\begin{split}
G^{ECEC}_{0\nu}=&\frac{1}{4R^2\ln2}\frac{(G\cos\theta)^4}{2\pi^2} (\hbar c^2)(m_ec^2)^7 \\
&\times{\cal B}_{n'_{e_1},-1}^2{\cal B}_{n'_{e_2},-1}^2 .
\end{split}
\end{equation}
The obtained prefactors are listed in Table~\ref{table2}.

Recently Krivoruchenko {\it et al.} \cite{kri11} have presented  a theory of PF slightly different from ours. The PF extracted from their paper is given in the last column of Table \ref{table2} for comparison. Note the large value of $G^{ECEC}_{0\nu}$ in both calculations for $^{180}$W decay, due to the large value of $Z\alpha=74/137\sim0.5$. For these heavy nuclei the PF is very sensitive to the treatment of the electron wave function near the origin $r=0$. Differences between the two PF calculations may in part arise from the way in which the nuclear size and electron screening correction is taken into account.

\begin{ruledtabular}
\begin{center}
\begin{table}[cbt!]
\label{table2}
\caption{Prefactors for neutrinoless double electron capture.}
\begin{tabular}{lcc}
Decay  &\multicolumn{2}{c}{$G^{ECEC}_{0\nu}(10^{-19} yr^{-1})$}\\ \cline{2-3}
       &This work &Ref. \cite{kri11} \\
\hline
\T
$_{54}^{124}$Xe$_{70}\rightarrow _{52}^{124}$Te$_{72}^*$ &2.57 &\\
$_{64}^{152}$Gd$_{88}\rightarrow _{62}^{152}$Sm$_{90}$ &1.46 &1.67\\
$_{66}^{156}$Dy$_{90}\rightarrow _{64}^{156}$Gd$_{92}^*$ &0.266 &0.22\\
$_{68}^{164}$Er$_{96}\rightarrow _{66}^{164}$Dy$_{98}$ &0.362 &0.31\\
$_{74}^{180}$W$_{106}\rightarrow _{72}^{180}$Hf$_{108}$ &46.2 &34.9\\
\end{tabular}
\end{table}
\end{center}
\end{ruledtabular}

\section{Nuclear matrix elements}
\begin{figure*}[cbt!]
\begin{center}
\includegraphics[width=14.cm]{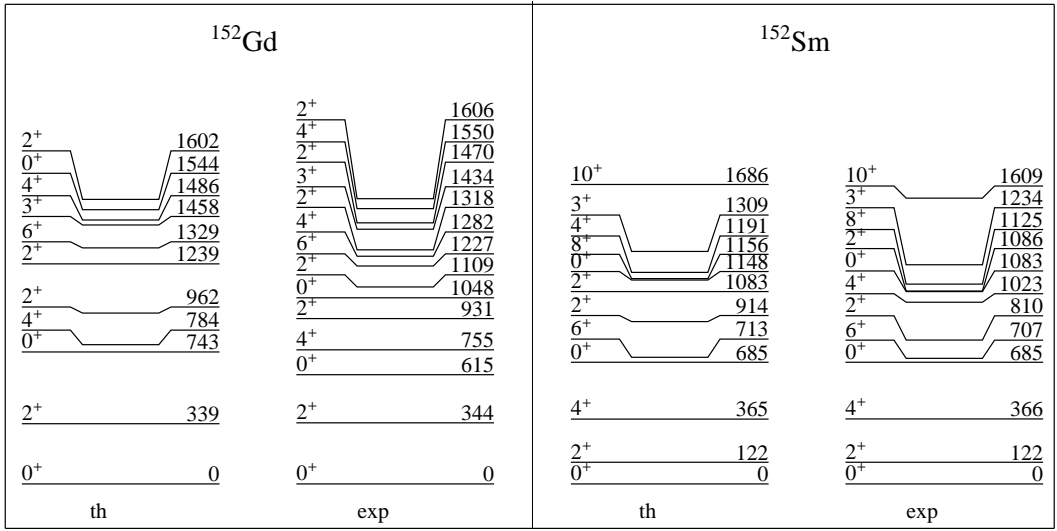} 
\end{center}
\caption{\label{fig152}Comparison between calculated and experimental low-lying spectra for $^{152}$Gd$\to ^{152}$Sm.}
\end{figure*}

\begin{figure*}[cbt!]
\begin{center}
\includegraphics[width=14.cm]{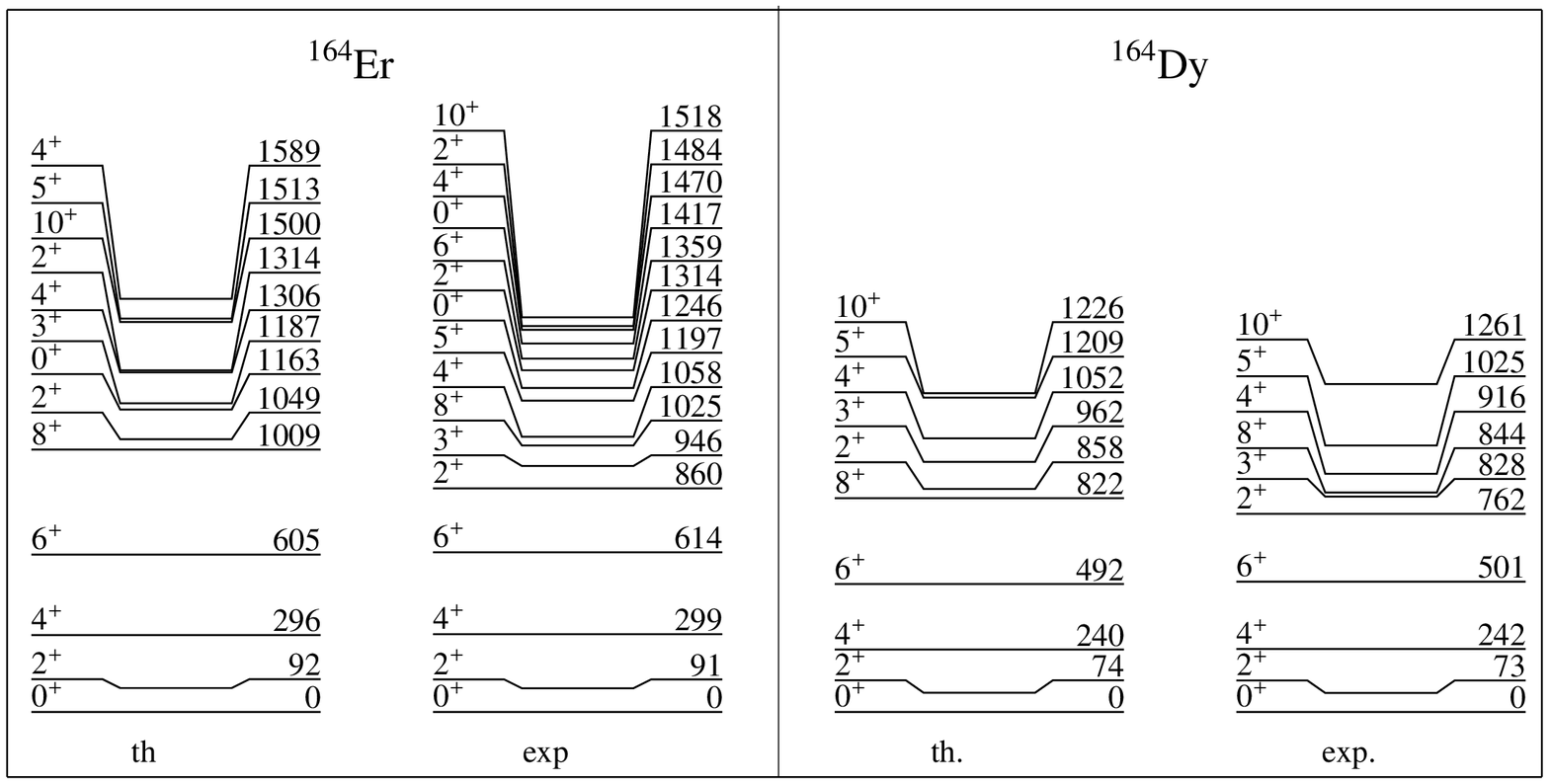} 
\end{center}
\caption{\label{fig164}Comparison between calculated and experimental low-lying spectra for $^{164}$Er$\to ^{164}$Dy.}
\end{figure*}

\begin{figure*}[cbt!]
\begin{center}
\includegraphics[width=14.cm]{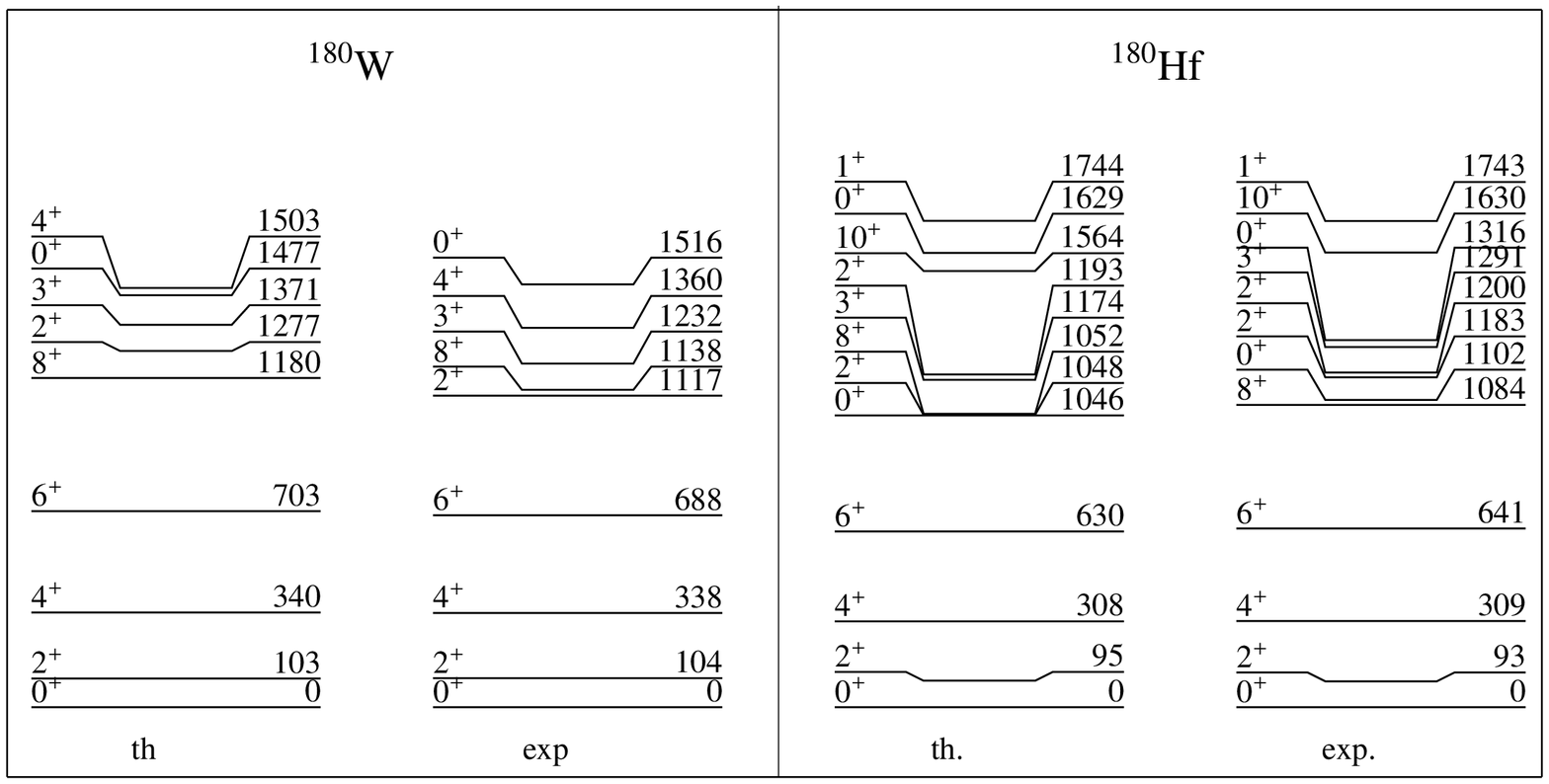} 
\end{center}
\caption{\label{fig180}Comparison between calculated and experimental low-lying spectra for $^{180}$W$\to ^{180}$Hf.}
\end{figure*}

The calculation of NME for the $0\nu ECEC$ is more difficult than for $0\nu \beta^- \beta^-$ because of two reasons: (i) In two cases $^{124}$Xe and $^{156}$Dy the resonant state is an excited state. (ii) In the other cases, the decay $^{152}$Gd is to a transitional nucleus, and the decays $^{164}$Er and $^{180}$W are to strongly deformed nuclei. For these reasons, one needs a model that can calculate reliably energies and wave functions of ground and excited states in spherical, transitional, and deformed nuclei. To this end we make use of the microscopic
interacting boson model (IBM-2) \cite{iac11}.
 The method is described  in Refs. \cite{bar09, bar12b}. We write
\begin{equation}
\begin{split}
M_{0\nu} & =  g_{A}^{2}M^{(0\nu)}, \\
M^{(0\nu)} & =  M_{GT}^{(0\nu)} -\left(\frac{g_{V}}{g_{A}}\right)^2 M_{F}^{(0\nu)}+M_{T}^{(0\nu)},
\end{split}
\end{equation}
with $M_{GT}^{(0\nu)}$, $M_{F}^{(0\nu)}$, and $M_{T}^{(0\nu)}$ defined as in Eq. (20) of Ref. \cite{bar12b}
\begin{equation}
\begin{split}
M^{(0\nu)}_{GT}\equiv\left\langle ^{A}\text{X;}0_{1}^{+}\left\vert
h^{GT}(p)/g^2_A \right\vert ^{A}\text{Y;}J_{F}\right\rangle, \\ 
M^{(0\nu)}_F\equiv\left\langle ^{A}\text{X;}0_{1}^{+}\left\vert
 h^{F}(p)/g^2_V \right\vert ^{A}\text{Y;}J_{F}\right\rangle ,\\
M^{(0\nu)}_T\equiv\left\langle ^{A}\text{X;}0_{1}^{+}\left\vert 
h^{T}(p)/g^2_A \right\vert ^{A}\text{Y;}J_{F}\right\rangle ,
\end{split}
\end{equation}
where $h(p)$ are the form factors tabulated in Table II of \cite{bar12b}. The wave functions of the initial and final states are taken either from the literature or from a fit to the observed energies and other properties ($B(E2)$ values, quadrupole moments, $B(M1)$ values, magnetic moments, etc.).
The values of the parameters used in the calculation are given in Appendix A. The quality of the wave functions as well as the quality of the description of energies and electromagnetic transition rates is excellent both in transitional A=152, Fig.~\ref{fig152}, and strongly deformed A=164, Fig.~\ref{fig164}, and A=180, Fig.~\ref{fig180}, nuclei.

\begin{figure*}[cbt!]
\begin{center}
\includegraphics[width=14.cm]{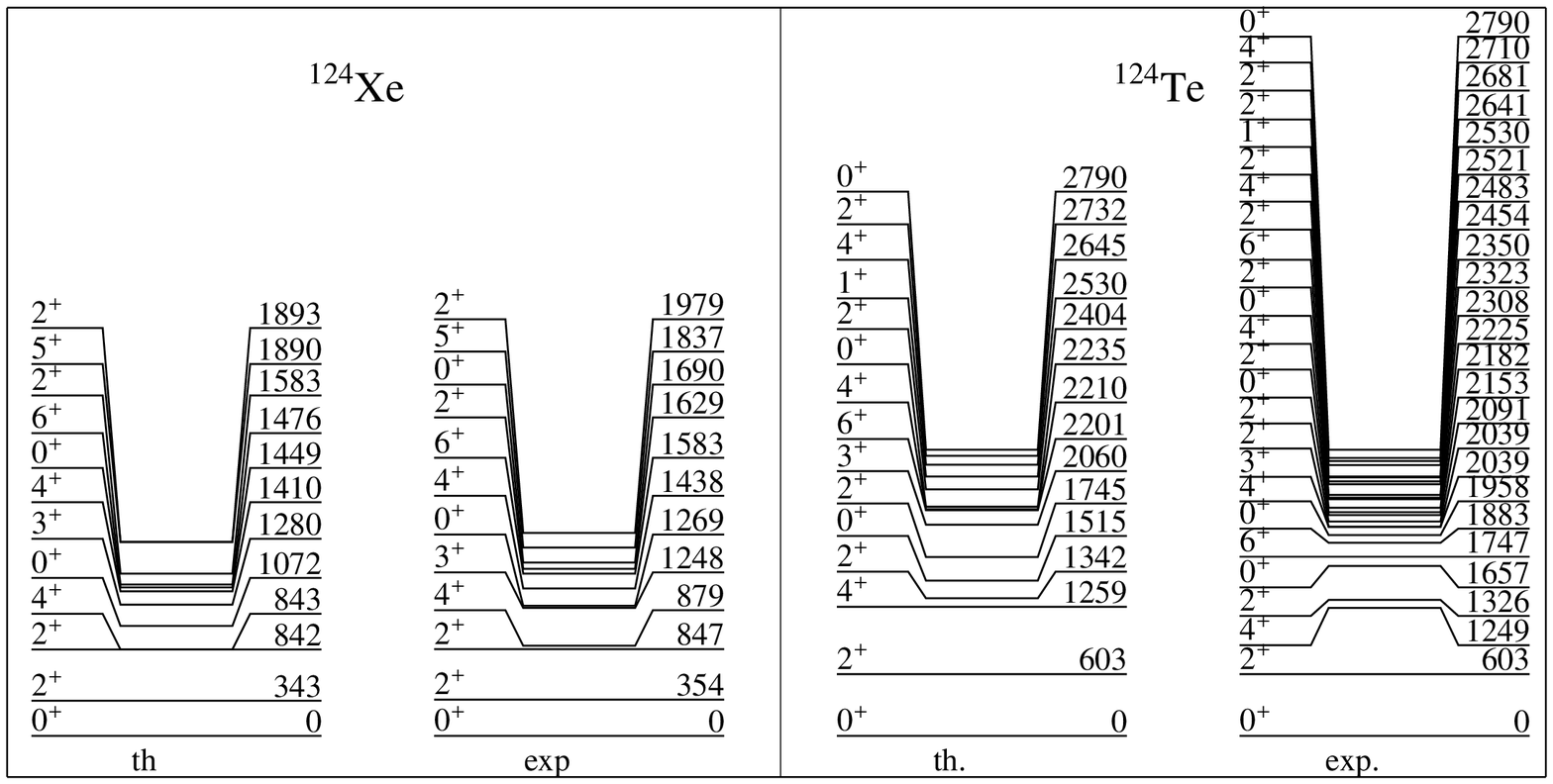} 
\end{center}
\caption{\label{fig124}Comparison between calculated and experimental low-lying spectra for $^{124}$Xe$\to ^{124}$Te.}
\end{figure*}

\begin{figure*}[cbt!]
\begin{center}
\includegraphics[width=14.cm]{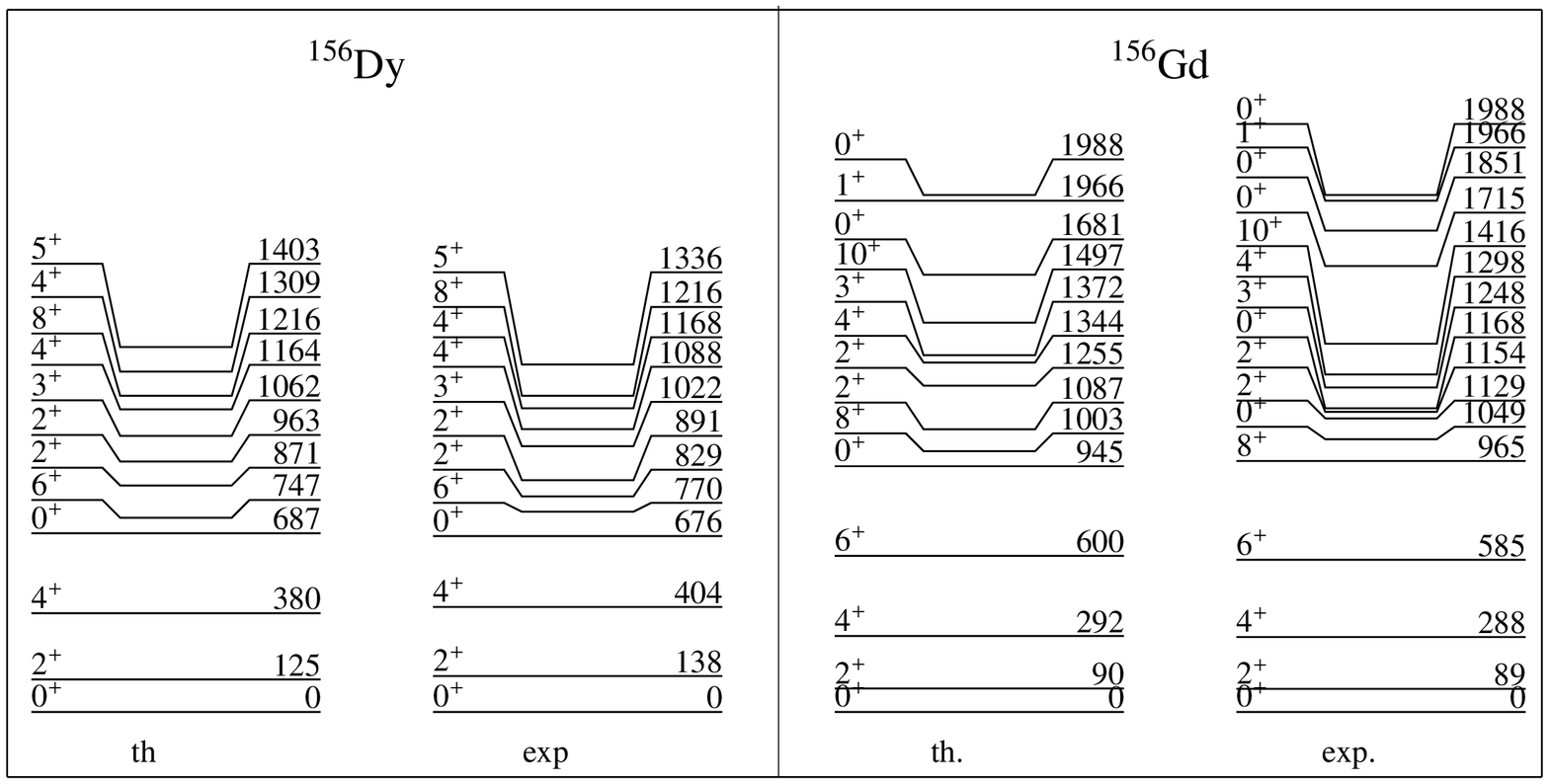} 
\end{center}
\caption{\label{fig156}Comparison between calculated and experimental low-lying spectra for $^{156}$Dy$\to ^{156}$Gd.}
\end{figure*}

For decays to excited states, particular care must be taken in identifying the state in the calculated spectrum. For decay to $^{156}$Gd this state is the $0^+_4$ state calculated at 1988 keV, while for decay to $^{124}$Te this state is also the $0^+_4$ state calculated at 2790 keV. The quality of the excited spectrum of states in  $^{124}$Te, Fig.~\ref{fig124}, and $^{156}$Gd, Fig.~\ref{fig156} is also excellent. 

Using IBM-2 wave functions and the theory previously described  \cite{bar09, bar12b} we can calculate the NME for neutrinoless ECEC decay shown in  Table \ref{table3}.
\begin{ruledtabular}
\begin{table*}[cbt!]
\caption{\label{table3}IBM-2 nuclear matrix elements $M^{(0\nu)}$ (dimensionless) for neutrinoless $ECEC$ decay with  Argonne SRC and $g_V/g_A=1/1.269$.}
\begin{tabular}{lcccccccc}
 & \multicolumn{4}{c}{$M^{(0\nu)}$ light} & \multicolumn{4}{c}{$M^{(0\nu)}_h$ heavy}\tabularnewline \cline{2-5} \cline{6-9}
\T
& $M_{GT}^{(0\nu)}$ & $M_{F}^{(0\nu)}$  & $M_{T}^{(0\nu)}$ & $M^{(0\nu)}$ & $M_{GT}^{(0\nu)}$ & $M_{F}^{(0\nu)}$  & $M_{T}^{(0\nu)}$ & $M_h^{(0\nu)}$ \tabularnewline
\hline 
\T
$^{124}$Xe$_\rightarrow ^{124}$Te$^* [0^+_4]$ 	&0.277	&-0.051	&-0.012	&0.297	&6.447&-2.921&-1.521&6.740\\
$^{152}$Gd$\rightarrow ^{152}$Sm$ [0^+_1]$ 	&2.132	&-0.352	&0.095	&2.445	&68.910&-31.400&12.810&101.200\\
$^{156}$Dy$\rightarrow ^{156}$Gd$^* [0^+_4]$ 	&0.265	&-0.048	&0.017	&0.311	&10.530&-4.739&2.616&16.090\\
$^{164}$Er$\rightarrow ^{164}$Dy$ [0^+_1]$ 	&3.456	&-0.444	&0.221	&3.952	&107.900&-46.820&32.870&169.800\\
$^{180}$W$\rightarrow ^{180}$Hf$ [0^+_1]$ 	&4.117	&-0.566	&0.204	&4.672	&118.700&-53.320&28.200&170.900\\
\end{tabular}
\end{table*}
\end{ruledtabular}
The values of the nuclear matrix elements to the excited states are considerably smaller than those to the ground state, due to the very different nature of these states. To illustrate this point we show in Table \ref{tablenew} the results of the calculation for $A=156$ leading to the first five $0^+$ states in $^{156}$Gd. We see that the matrix elements of $0^+_3$, $0^+_4$, and $0^+_5$ are one order of magnitude smaller than $0^+_1$ and $0^+_2$.
This has a major consequence on $0\nu ECEC$ essentially excluding as possible candidates all decays to excited states. A similar conclusion was drawn in Ref.~\cite{suh12} for the decay $^{96}$Ru$\to ^{96}$Mo$^*$.

\begin{ruledtabular}
\begin{table*}[cbt!]
\caption{\label{tablenew}IBM-2 nuclear matrix elements $M^{(0\nu)}$ (dimensionless) for $A=156$ neutrinoless $ECEC$ decay to the first five $0^+$ states with  Argonne SRC and $g_V/g_A=1/1.269$ .}
\begin{tabular}{lcccccccc}
 & \multicolumn{4}{c}{$M^{(0\nu)}$ light} & \multicolumn{4}{c}{$M^{(0\nu)}_h$ heavy}\tabularnewline \cline{2-5} \cline{6-9}
\T
& $M_{GT}^{(0\nu)}$ & $M_{F}^{(0\nu)}$  & $M_{T}^{(0\nu)}$ & $M^{(0\nu)}$ & $M_{GT}^{(0\nu)}$ & $M_{F}^{(0\nu)}$  & $M_{T}^{(0\nu)}$ & $M_h^{(0\nu)}$ \tabularnewline
\hline 
\T
$0^+_1$ &2.796 &-0.398&0.132&3.175 	&82.560&-36.990&17.480&123.000\\
$0^+_2$ &1.532 &-0.227&0.076&1.749 	&47.630&-21.360&10.430&71.320\\
$0^+_3$ &0.403 &-0.065&0.022&0.466 	&13.900&-6.242&3.225&21.000\\
$0^+_4$ &0.265 &-0.048&0.017&0.311 	&10.530&-4.739&2.616&16.090\\
$0^+_5$ &0.302 &-0.046&0.016&0.346 	&9.809&-4.402&2.204&14.750\\
\end{tabular}
\end{table*}
\end{ruledtabular}

NME to the g.s of $^{152}$Sm, $^{164}$Dy, and $^{180}$Hf for light neutrino exchange have also been calculated in a variety of methods. In Table \ref{tablecomparison} we compare our results to these other calculations. While the QRPA matrix elements are of the same order of magnitude as IBM-2, the EDF results are much smaller, in particular they show a large reduction for the deformed nuclei $^{164}$Dy, and $^{180}$Hf. The origin of this discrepancy between IBM-2/QRPA and EDF is not clear. As shown in Fig.~\ref{fig180} the quality of the spectrum (and the electromagnetic transitions not shown) in IBM-2 is excellent including the location of the $\beta$ and $\gamma$ bands, which is crucial for providing good wave functions of collective states. Also the collective structure of the parent nucleus $^{180}$W and daughter nucleus  $^{180}$Hf is practically identical, with a large overlap, and both nuclei have many collective pairs, $N_{\pi}=$4 or 5 and $N_{\nu}=$10 or 9, respectively, that contribute to the decay. Although there is a reduction in the NME due to the deformation, we still would expect the matrix elements to be comparable to those in spherical nuclei.
\begin{ruledtabular}
\begin{center}
\begin{table}[cbt!]
\caption{\label{tablecomparison}
Comparison between IBM-2 matrix elements with Argonne SRC for $0\nu ECEC$ decay and QRPA and EDF.}
\begin{tabular}{lcccc}
Decay  &\multicolumn{3}{c}{$M^{(0\nu)}$ (light)}\\ \cline{2-4}
       & &spherical &deformed & \\ 
       &IBM-2 &QRPA\footnotemark[1] &QRPA\footnotemark[1] &EDF\footnotemark[2] \\ 
\hline
\T
$_{64}^{152}$Gd$_{88}\rightarrow _{62}^{152}$Sm$_{90}$ &2.44&7.59&3.23-2.67&1.07-0.89\\
$_{68}^{164}$Er$_{96}\rightarrow _{66}^{164}$Dy$_{98}$  &3.95&6.12&2.64-1.79&0.64-0.50\\
$_{74}^{180}$W$_{106}\rightarrow _{72}^{180}$Hf$_{108}$ &4.67&5.79&2.05-1.79&0.58-0.38\\
\end{tabular}
\footnotetext[1]{Ref.~\cite{fan12}}
\footnotetext[2]{Ref.~\cite{rod12}}
\end{table}
\end{center}
\end{ruledtabular}

\section{Half-lives}
\begin{figure}[cbt!]
\begin{center}
\includegraphics[width=8.6cm]{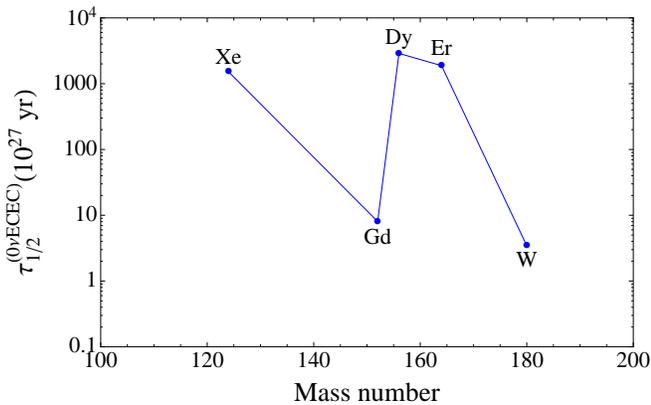} 
\end{center}
\caption{\label{fig5} (Color online) 
Expected half-lives for light neutrino exchange with $\left\langle m_{\nu}\right\rangle=1$~eV, $g_{A}=1.269$. The figure is in semilogarithmic scale.}
\end{figure}

\begin{figure}[cbt!]
\begin{center}
\includegraphics[width=8.6cm]{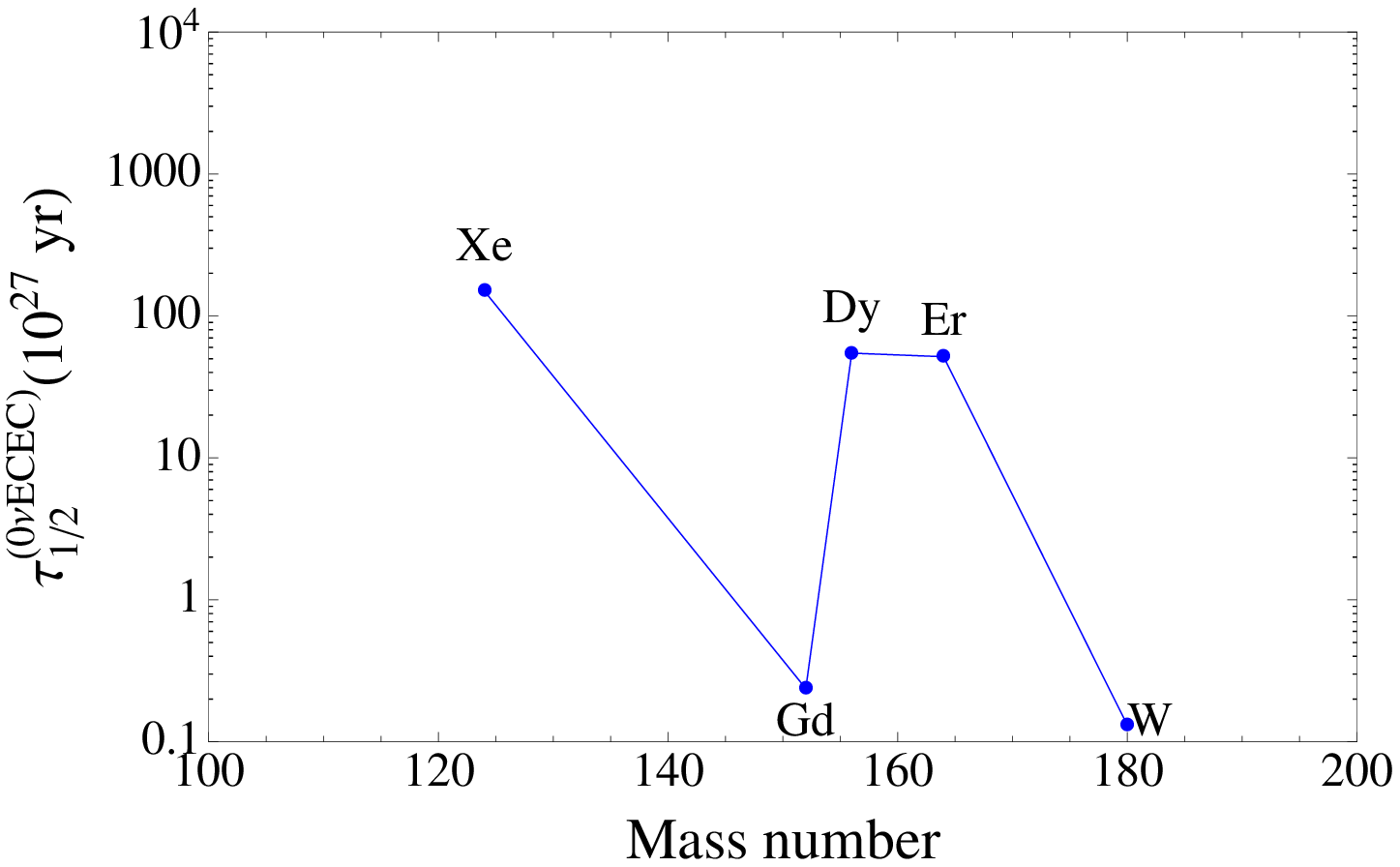} 
\end{center}
\caption{\label{fignew3} (Color online) 
Expected half-lives for heavy neutrino exchange with $m_{p}\left\langle m_{\nu_{h}}^{-1}\right\rangle =2.75\times10^{-7}$, $g_{A}=1.269$. The figure is in semilogarithmic scale.}
\end{figure}

The results for PF and NME of the previous sections can be combined to give the  half-lives in Table \ref{table4}  and Figs. \ref{fig5} and \ref{fignew3}.
The best case scenarios appear to be $^{152}$Gd$\rightarrow ^{152}$Sm$ [0^+_1]$ and
$^{180}$W$\rightarrow ^{180}$Hf$ [0^+_1]$, for which, however, the predicted half-life even with unquenched value of $g_A=1.269$ is the order of $10^{27}$yr. 

As in the previous papers,  an important question is the quenching of the $g_A$ which occurs to the fourth power in Eq. (2). To estimate this effect we use the parametrization of Eq. (40) of Ref. \cite{bar12b}, (maximal quenching)

\begin{equation}
g_{A, eff}^{\rm{IBM-2}}=1.269A^{-0.18}.
\end{equation}
With this parametrization we obtain the values of Table \ref{table4} (right).
\begin{ruledtabular}
\begin{table}[h]
\caption{\label{table4}Left: Calculated half-lives in IBM-2 for neutrinoless double electron capture with light neutrino exchange for $\left<m_{\nu}\right>=1$~eV   and heavy neutrino exchange with $m_{p}\left\langle m_{\nu_{h}}^{-1}\right\rangle =2.75\times10^{-7}$, and using unquenched value $g_A=1.269$. Right: Same as right but using maximally quenched value $g_{A, eff}^{\rm{IBM-2}}=1.269A^{-0.18}$.}
\begin{tabular}{lcccc}
 &\multicolumn{4}{c}{$\tau_{1/2}(10^{27})$yr}\\\cline{2-5}
 &\multicolumn{2}{c}{unquenched}  &\multicolumn{2}{c}{maximally quenched} \\ \cline{2-3} \cline{4-5}
Nucleus&light &heavy &light &heavy\\
\hline
\T
$^{124}$Xe$_\rightarrow ^{124}$Te$^* [0^+_4]$ &1520	&150 &49000  &4820\\
$^{152}$Gd$\rightarrow ^{152}$Sm$ [0^+_1]$ 	&8.03	&0.237 &298 &8.83\\
$^{156}$Dy$\rightarrow ^{156}$Gd$^* [0^+_4]$ &2890	&54.7 &110000 &	2080\\
$^{164}$Er$\rightarrow ^{164}$Dy$ [0^+_1]$ 	&1880	&51.6	&74000&2030\\
$^{180}$W$\rightarrow ^{180}$Hf$ [0^+_1]$ 	&3.44	&0.131	&144&5.47\\
\end{tabular}
\end{table}
\end{ruledtabular}

\section{Conclusions}
In this paper, we have presented both PFs and NMEs for neutrinoless double electron capture and from these calculated the expected half-lives for both light and heavy neutrino exchange. The values obtained are longer than those for $0\nu\beta^-\beta^-$ decay. The best cases are those of $^{180}$W$\rightarrow ^{180}$Hf$ [0^+_1]$ and $^{152}$Gd$\rightarrow ^{152}$Sm$ [0^+_1]$, where for light neutrino exchange $\tau_{1/2}=3.44\times 10^{27}$yr and 8.03 $\times 10^{27}$yr, respectively, for $\left<m_{\nu}\right>=1$~eV and $g_A=1.269$. For comparison $0\nu\beta^-\beta^-$ decays have $\tau_{1/2}$ of the order  of $10^{24}$yr. The half-lives for $0\nu ECEC$ are, however of the same order of magnitude of $0\nu\beta^+\beta^+$, the best case for this decay being that of  $^{124}$Xe$\rightarrow ^{124}$Te which has $\tau_{1/2}=3.32 \times10^{27}$yr. Our conclusion is that, even in the optimistic scenario of $g_A=1.269$, $0\nu ECEC$ is unreachable in the present generation of experiments.

\section{Acknowledgements}
This work was supported in part by US Department of Energy Grant no. DE-FG-02-91ER-40608, Fondecyt Grant No. 1120462, and Academy of Finland Grant No. 266437.

\section{Appendix A: Parameters of the IBM-2 Hamiltonian}

A detailed description of the IBM-2 Hamiltonian is given in \cite{iac1} and \cite{otsukacode}. For $^{124}\mbox{Xe}$, $^{152}\mbox{Gd}$, $^{156}\mbox{Dy}$, $^{156}\mbox{Gd}$, and $^{164}\mbox{Gd}$ the Hamiltonian parameters are taken from the literature.  The new calculations are done using the program NPBOS \cite{otsukacode} adapted by J. Kotila. The values of the Hamiltonian parameters, as well as the references
from which they are taken, are given in Table~\ref{tab:ibm2parameters}.

\begin{table*}[ctb!]
 \caption{Hamiltonian parameters employed in the IBM-2 calculation along with their references.}
\label{tab:ibm2parameters}
\begin{ruledtabular}
\begin{centering}
\begin{tabular}{lccccccccccccccccccc}
Nucleus  & $\epsilon_{d_{\nu}} = \epsilon_{d_{\pi}}$  & $\kappa$  & $\chi_{\nu}$  & $\chi_{\pi}$  & $\xi_{1}$  & $\xi_{2}$  & $\xi_{3}$  & $c_{\nu}^{(0)}$  & $c_{\nu}^{(2)}$  & $c_{\nu}^{(4)}$  & $c_{\pi}^{(0)}$  & $c_{\pi}^{(2)}$  & $c_{\pi}^{(4)}$  & $\omega_{\nu\nu}$  & $\omega_{\pi\pi}$  & $\omega_{\nu\pi}$ \tabularnewline
\hline 
\T
$^{124}\mbox{Xe}$ \cite{Puddu80}  & 0.70   & -0.14  & 0.00  & -0.80  & -0.18  & 0.24  & -0.18  &0.05  &-0.16  &  &  &  &  &  &  &   \tabularnewline
$^{124}\mbox{Te}$ \footnotemark[1]   & 0.83  & -0.15  & 0.00  & -1.20  &-0.34  & 0.15  & -0.34  &0.10  &  &  &  &  &  &   &  & \tabularnewline
$^{152}\mbox{Gd}$ \cite{kot12d}  & 0.74   & -0.076  & -0.80  & -1.00  &0.08  &0.08  &0.08  &  &  &  &-0.20  &0.10  &  &  &  &   \tabularnewline
$^{152}\mbox{Sm}$\footnotemark[1]  & 0.52  & -0.075  & -1.00  & -1.30  &0.3  &-0.01  & 0.03  &  &  &  &  &0.05  &  &  &  &  \tabularnewline
$^{156}\mbox{Dy}$ \cite{kot12d}  & 0.62 & -0.078  &-1.00  & -0.80  & 0.08  &0.08  & 0.08  &  &  &  &-0.05   &-0.15 &  &  &  &  \tabularnewline
$^{156}\mbox{Gd}$ \cite{kot12d}  & 0.59   & -0.08  & -1.10  & -1.00  &0.11  &0.09  &0.11  &  &  &  &-0.20  &-0.10  &  & -0.0025  & -0.0025  & -0.0025  \tabularnewline
$^{164}\mbox{Er}$ \footnotemark[1]  & 0.47    & -0.08  & -0.50  & 0.70  & 0.24  & 0.24  & 0.24  & -0.28  &   &  &  &  &  &  &  &  \tabularnewline
$^{164}\mbox{Dy}$ \cite{kot12d}  & 0.54    & -0.05  & -0.70  &-0.80   & 0.17  & 0.17  &0.17   &   &  &  & &-0.15  &  &  &  &   \tabularnewline
$^{180}\mbox{W}$ \footnotemark[1]  & 0.53   & -0.11  & -0.10  & -1.60  &-0.02  &0.04  &-0.02  & -0.11  &   &   &  &  &  &  &  &  &  & \tabularnewline
$^{180}\mbox{Hf}$ \footnotemark[1]  & 0.53   & -0.08  & -0.3  & -1.20  & 0.05  & 0.05  & 0.05  & -0.14  &   &   &   &   &   &  &  &   \tabularnewline
\tabularnewline
\end{tabular}
\footnotetext[1]{Parameters fitted to reproduce the spectroscopic
data.}
\end{centering}
 \end{ruledtabular} 
\end{table*}

\end{document}